\documentclass[]{aastex631}

\usepackage{amsmath}
\usepackage{booktabs}

\begin{document}

\title{Identification of the weak-to-strong transition in Alfvénic turbulence from space plasma}

\author[0000-0003-4268-7763]{Siqi Zhao}
\affiliation{Deutsches Elektronen Synchrotron DESY, Platanenallee 6, D-15738, Zeuthen, Germany}
\affiliation{Institut für Physik und Astronomie, Universität Potsdam, D-14476, Potsdam, Germany}

\author[0000-0003-2560-8066]{Huirong Yan}
\affiliation{Deutsches Elektronen Synchrotron DESY, Platanenallee 6, D-15738, Zeuthen, Germany}
\affiliation{Institut für Physik und Astronomie, Universität Potsdam, D-14476, Potsdam, Germany}

\author[0000-0003-1778-4289]{Terry Z. Liu}
\affiliation{Department of Earth, Planetary, and Space Sciences, University of California, Los Angeles, CA 90024, USA}
 \email{huirong.yan@desy.de; terryliuzixu@ucla.edu}

\author[0000-0003-1683-9153]{Ka Ho Yuen}
\affiliation{Theoretical Division, Los Alamos National Laboratory, Los Alamos, NM 87545, USA}

\author[0000-0002-3652-6210]{Huizi Wang}
\affiliation{Shandong Key Laboratory of Optical Astronomy and Solar-Terrestrial Environment, Institute of Space Sciences, 264209, Shandong University, Weihai, People’s Republic of China}


\begin{abstract}

Plasma turbulence is a ubiquitous dynamical process that transfers energy across many spatial and temporal scales in astrophysical and space plasma systems. Although the theory of anisotropic magnetohydrodynamic (MHD) turbulence has successfully described natural phenomena, its core prediction of an Alfvénic transition from weak to strong MHD turbulence when energy cascades from large to small scales has not been observationally confirmed. Here we report evidence for the Alfvénic weak-to-strong transition in small-amplitude, turbulent MHD fluctuations in Earth's magnetosheath using data from four \textit{Cluster} spacecraft. Our observations demonstrate the universal existence of strong turbulence accompanied by weak turbulent fluctuations on large scales. Moreover, we find that the nonlinear interactions of MHD turbulence are crucial to the energy cascade, broadening the cascade directions and fluctuating frequencies. The observed connection between weak and strong MHD turbulence systems can be present in many astrophysical environments, such as star formation, energetic particle transport, turbulent dynamo, and solar corona or solar wind heating.

\end{abstract}



\section{Introduction} \label{sec:intro}

The theory of anisotropic MHD turbulence has been widely accepted and adopted in plasma systems, ranging from clusters of galaxies, the interstellar medium, accretion disks, to the heliosphere \cite{Yan2002,Brunetti2007,Bruno2013}. One of the most crucial predictions of the theory is an Alfvénic transition from weak to strong MHD turbulence when energy cascades from large to small scales \cite{Goldreich1995,Howes2011}. The self-organized process from weak to strong MHD turbulence is the cornerstone of understanding the energy cascade in the complete picture of MHD turbulence. 

The critical balance model is an attractive model for describing physical behaviors of incompressible MHD (IMHD) turbulence \cite{Goldreich1995}. When $\tau_A \ll \tau_{nl}$ (referred to as weak MHD turbulence), weak interactions among the counter-propagating wave packets transfer energy to higher $k_\perp$, whereas no energy cascades to higher $k_\parallel$, where $\tau_A=1/(k_\parallel V_A)$ is the linear Alfvén wave time, $\tau_{nl}=1/(k_\perp \delta V_\perp)$ is the nonlinear time, $V_A$ is the Alfvén speed, $\delta V_\perp$ is the fluctuating velocity perpendicular to the background magnetic field ($\mathbf{B}_0$), and $k_\perp$ and $k_\parallel$ are wavenumbers perpendicular and parallel to $\mathbf{B}_0$ \cite{Galtier2000}. As turbulence cascades to smaller scales, nonlinearity strengthens until reaching the critical balance $(\tau_A\approx\tau_{nl})$ at the transition scale ($\lambda_{CB}$). On scales smaller than $\lambda_{CB}$, Alfvén wave packets are statistically destroyed in one $\tau_A$. In addition to the first order interactions of counter-propagating waves, all higher orders of interactions can contribute, creating strong MHD turbulence \cite{Goldreich1995,Mallet2015}. In compressible MHD (CMHD), small-amplitude fluctuations can be decomposed into three eigenmodes (namely, Alfvén, fast, and slow modes) in homogeneous plasma \cite{Cho2003,Makwana2020,Zhu2020,Chaston2020,Zhao2021}. Alfvén modes decoupled from CMHD turbulence, linearly independent of fast and slow modes, show similar properties to those in IMHD turbulence, e.g., the Kolmogorov spectrum and the scale-dependent anisotropy \cite{Cho2003,Zhao2022}. Additionally, numerical simulations have confirmed that the Alfvénic weak-to-strong transition occurs in both IMHD turbulence and Alfvén modes decomposed from CMHD turbulence \cite{Verdini2012,Meyrand2016,Makwana2020}.

However, such a transition has not been confirmed from observations. In this study, we present evidence for the Alfvénic weak-to-strong transition and estimate the transition scale $\lambda_{CB}$ in Earth's magnetosheath using data from four \textit{Cluster} spacecraft \cite{Escoubet2001}. Earth's magnetosheath offers a representative environment for studying plasma turbulence, given that most astrophysical and space plasmas with finite plasma $\beta$ are compressible, where $\beta$ is the ratio of the plasma to magnetic pressure.

\section{Results} \label{sec:floats}

Here we present an overview of fluctuations observed by \textit{Cluster-1} in geocentric-solar-ecliptic (GSE) coordinates during 23:00-10:00 Universal Time (UT) on 2-3 December in Fig. 1. During this period, four \textit{Cluster} spacecraft flew in a tetrahedral formation with relative separation $d_{sc}\approx200km$ (around 3 proton inertial length $d_{i}\approx74km$) on the flank of Earth's magnetosheath around [1.2, 18.2, -5.7] $R_E$ (Earth radius). We choose this time interval to study the Alfvénic weak-to-strong transition because the fluctuations satisfy the following criteria. Firstly, the background magnetic field ($\mathbf{B}$) measured by the Fluxgate Magnetometer (FGM) \cite{Balogh1997} and the proton bulk velocity ($\mathbf{V}_p$) measured by the Cluster Ion Spectrometry (CIS) \cite{Reme2001} are relatively stable in Figs. 1a-b. We cross-verify the reliability of plasma data, based on the consistency between the proton density ($N_p$) measured by CIS and electron density measured by the Waves of High frequency and Sounder for Probing of Electron density by Relaxation (WHISPER) \cite{Decreau1997} in Fig. 1c. 

We set a moving time window with a five-hour length and a five-minute moving step. The selection of a five-hour length ensures that we obtain measurements at low frequencies (large scales) while the mean magnetic field ($\mathbf{B}_0$) within the moving time window is approaching the local mean field at the selected largest spatial scale. The uniform $\mathbf{B}_0$ is independent of the transformation between real and wavevector space; however, it differs from the theoretically expected local mean field at each scale. To assess such differences, Supplementary Fig. 1 shows that the local mean field at different scales is closely aligned with $\mathbf{B}_0$ most of the time, suggesting that $\mathbf{B}_0$ approximating the local mean field is acceptable. To further address this limitation of the mode decomposition method, which relies on a perturbative treatment of fluctuations in the presence of a uniform background magnetic field \cite{Cho2003}, we provide results obtained using various time window lengths, all of which show similar conclusions (Supplementary Fig. 2).

Fig. 1d shows spectral slopes of the trace magnetic field and proton velocity power calculated by fast Fourier transform (FFT) with three-point centered smoothing in each time window. These spectral slopes at spacecraft-frame frequency $f_{sc}\approx [0.001Hz,0.1f_{ci}]$ are close to $-5/3$ or $-3/2$ (the proton gyro-frequency $f_{ci}\approx0.24Hz$), suggesting turbulent fluctuations are in a fully-developed state. The remaining magnetosheath fluctuations with spectra close to $f_{sc}^{-1}$ are typically populated by uncorrelated fluctuations \cite{Hadid2015,Hadid2018} and are beyond the scope of the present paper. Fig. 1e shows the average proton plasma $\beta_p$ is around $1.4$. Finally, Fig. 1f shows that the turbulent Alfvén number $M_{A,turb}\equiv\delta V_p/V_A\approx \delta B/(2B_0)\approx 0.33$, suggesting that fluctuations include substantial Alfvénic components and satisfy the small-amplitude fluctuation assumption (the nonlinear terms ($\delta V_p^2$, $\delta B^2$) are weaker than the linear terms ($V_A\delta V_p$, $B_0\delta B$)). Nevertheless, the average magnetic compressibility $C_\parallel(f_{sc})=\frac{|\delta B_\parallel(f_{sc})|^2}{|\delta B_\parallel(f_{sc})|^2 + |\delta B_\perp(f_{sc})|^2}\approx 0.34$, indicating fluctuations are a mixture of Alfvén and compressible magnetosonic (fast and slow) modes \cite{Sahraoui2020}, where $\delta B_\parallel$ and $\delta B_\perp$ are the fluctuating magnetic field parallel and perpendicular to $\mathbf{B}_0$.

Due to the homogeneous and stationary state of the turbulence (Supplementary Fig. 3), we can utilize frequency-wavenumber distributions of Alfvénic power, i.e. magnetic power $P_{B_A}(k_\perp,k_\parallel,f_{sc})$ and proton velocity power $P_{V_A}(k_\perp,k_\parallel,f_{sc})$, to investigate the structure of turbulence. Alfvénic fluctuations are extracted based on their incompressibility and fluctuating directions perpendicular to $\mathbf{B}_0$ (see Methods). The extraction is taken at each time window. To distinguish spatial and temporal evolutions without any spatiotemporal hypothesis, we determine wavevectors by combining the singular value decomposition method \cite{Santolik2003} (to obtain $\mathbf{\hat{k}}_{SVD}$) and multispacecraft timing analysis \cite{Pincon1998} (to obtain $\mathbf{\hat{k}}_A$). It is worth noting that $\mathbf{\hat{k}}_A$ is not completely aligned with $\mathbf{\hat{k}}_{SVD}$. Namely, $\mathbf{\hat{k}}_A$ may deviate from $\mathbf{\hat{k}}_{SVD}$ by angle $\eta$ (Supplementary Fig. 4). Thus, we present the results under $\eta<10^\circ$, $\eta<15^\circ$, $\eta<20^\circ$, $\eta<25^\circ$, and $\eta<30^\circ$. Given the marginal impact of different choices of $\eta$ (Supplementary Figs. 5 and 6), spectral results are displayed by taking the data set under $\eta<30^\circ$ as an example without loss of generality. The fluctuations, in this event, count for 42$\%$ of total Alfvénic fluctuations.

To ensure the reliability of wavenumber determination, we establish a minimum threshold of $k>1/(100d_{sc})$ and $k_\parallel>10^{-5} km^{-1}$. Consequently, our observations exclude the ideal two-dimensional (2D) case ($k_\parallel=0$), where $\tau_A$ is infinity, indicating persistent strong nonlinearity \cite{Galtier2003,Nazarenko2007}. Nevertheless, quasi-2D (small $k_\parallel$) modes are present, as $k_\parallel$ is much smaller than $k_\perp$ at small wavenumbers (Fig. 2). These quasi-2D fluctuations satisfies $\tau_A<\tau_{nl}$, as shown in Fig. 3c, and exhibit weak nonlinearity. This weak turbulent state occurs since $\delta B_A^2(k_\perp,k_\parallel)/B_0^2$ is very low, where Alfvénic magnetic energy density at $k_\perp$ and $k_\parallel$ is calculated by $\delta B_A^2(k_\perp,k_\parallel)=\sum_{k_\perp=k_\perp}^{k_\perp\rightarrow\infty}\sum_{k_\parallel=k_\parallel}^{k_\parallel\rightarrow\infty}\int_0^\infty P_{B_A}(k_\perp,k_\parallel,f_{sc})df_{sc}$. 

\noindent\textbf{Evidence for the Alfvénic weak-to-strong transition}

Two-dimensional wavenumber distributions of magnetic energy are calculated by
\begin{eqnarray}
    D_{B_A}(k_\perp,k_\parallel)=\int_0^\infty P_{B_A}(k_\perp,k_\parallel,f_{sc})df_{sc}
    \label{1}
\end{eqnarray}
$\hat{D}_{B_A}(k_\perp,k_\parallel)=D_{B_A}(k_\perp,k_\parallel)/D_{B_A,max}$ is normalized by the maximum magnetic energy in all $(k_\perp,k_\parallel)$ bins, displayed by the spectral image and contours in Fig. 2. Compared to the isotropic dotted curves, $\hat{D}_{B_A}(k_\perp,k_\parallel)$ is prominently distributed along the $k_\perp$ direction, suggesting a faster perpendicular cascade. This anisotropic behavior is more pronounced at higher wavenumbers, consistent with previous simulations and observations \cite{Cho2000,He2011,Makwana2020,Zhao2022}. 

Moreover, $\hat{D}_{B_A}(k_\perp,k_\parallel)$ is compared with  the modeled 2D theoretical energy spectra based on strong turbulence \cite{Yan2008,Goldreich1995}
\begin{eqnarray}
    I_A(k_\perp,k_\parallel)\propto k_\perp^{-7/3} \exp(-\frac{L_0^{1/3}|k_\parallel|}{M_{A,turb}^{4/3}k_\perp^{2/3}})
    \label{2},
\end{eqnarray}
where the injection scale $L_0\approx[4.6\times10^4,8.1\times10^4]km$ is approximately estimated by the correlation time $T_c\approx[1300,2300]s$ and rms perpendicular fluctuating velocity $\delta V_{p\perp}\approx M_{A,turb}V_A\approx35km s^{-1}$. In Fig. 2, $\hat{I}_{A}(k_\perp,k_\parallel)$ is normalized $I_{A}(k_\perp,k_\parallel)$ by a constant value (one-third of the maximum magnetic energy in all $(k_\perp,k_\parallel)$ bins), displayed by color contours with black dashed curves. The 2D distribution $\hat{D}_{B_A}(k_\perp,k_\parallel)$ shows two different properties: (1) At $k_\perp<2\times10^{-4}km^{-1}$, $\hat{D}_{B_A}(k_\perp,k_\parallel)$ is mainly concentrated at $k_\parallel<7\times10^{-5}km^{-1}$ cascading along $k_\perp$ direction, suggesting that little energy cascade parallel to the background magnetic field, consistent with energy distributions in weak MHD turbulence \cite{Galtier2000}. (2) At $k_\perp>2\times10^{-4}km^{-1}$, $\hat{D}_{B_A}(k_\perp,k_\parallel)$ starts to distribute to higher $k_\parallel$, and both wavenumber distributions and intensity changes of $\hat{D}_{B_A}(k_\perp,k_\parallel)$ are almost consistent with $\hat{I}_{A}(k_\perp,k_\parallel)$. It indicates that $\hat{D}_{B_A}(k_\perp,k_\parallel)$ captures some theoretical characteristics of strong MHD turbulence \cite{Goldreich1995}. Besides, $\hat{D}_{B_A}(k_\perp,k_\parallel)$ is in good agreement with the Goldreich-Sridhar scaling $k_\parallel\propto k_\perp^{2/3}$ \cite{Goldreich1995}. This result further confirms that the properties of $\hat{D}_{B_A}(k_\perp,k_\parallel)$ at $k_\perp>2\times10^{-4}km^{-1}$ are closer to those in strong MHD turbulence. The change in $\hat{D}_{B_A}(k_\perp,k_\parallel)$ from purely stretching along the $k_\perp$ direction to following the Goldreich-Sridhar scaling $k_\parallel\propto k_\perp^{2/3}$ reveals a possible transition in the energy cascade.

Fig. 3a shows the compensated spectra ($k_\perp^{5/3}E_{B_A}(k_\perp)$), where the magnetic energy spectral density is defined as $E_{B_A}(k_\perp)=\frac{\delta B_A^2(k_\perp)}{2k_\perp}$, and $\delta B_A^2(k_\perp)$ is magnetic energy density at $k_\perp$ (see Methods). In Zone (2), $k_\perp^{5/3}E_{B_A}(k_\perp)$ is roughly consistent with $k_\perp^{5/3-2}$ (the dashed line), indicating that spectral slopes of $E_{B_A}(k_\perp)$ are around $-2$. In Zone (3), on the other hand, $k_\perp^{5/3}E_{B_A}(k_\perp)$ is almost flat, suggesting that $E_{B_A}(k_\perp)$ satisfies the Kolmogorov scaling ($E_{B_A}(k_\perp)\propto k_\perp^{-5/3}$). The sharp change in spectral slopes of $E_{B_A}(k_\perp)$ from $-2$ to $-5/3$ is apparent evidence for the transition of turbulence regimes \cite{Verdini2012,Meyrand2016}. In addition, $E_{B_A}(k_\perp)\propto k_\perp^{-1}$ appears in a substantial portion of Zone (1), indicating the weak turbulence forcing in action \cite{Schekochihin2012,Makwana2020}.

Fig. 3b shows the variation of $k_\parallel$ versus $k_\perp$ given the same Alfvénic magnetic energy. As $k_\perp$ increases, $k_\parallel$ is relatively stable at $k_\parallel\approx7\times10^{-5} km^{-1}$ in Zone (1). In Zone (3), the variation of $k_\perp$ versus $k_\parallel$ agrees with the Goldreich-Sridhar scaling $k_\parallel\propto k_\perp^{2/3}$ (the dashed line). Fig. 3c shows $k_\perp-k_\parallel$ distributions of nonlinearity parameter ($\chi_{B_A}(k_\perp,k_\parallel)$), which is one of the most critical parameters in distinguishing between weak and strong MHD turbulence \cite{Howes2011}, where $\chi_{B_A}(k_\perp,k_\parallel)$ is calculated by $\frac{k_\perp \delta B_A(k_\perp,k_\parallel)}{k_\parallel B_0}$ (see Methods). At the corresponding parallel and perpendicular wavenumbers in Fig. 3b, $\chi_{B_A}(k_\perp,k_\parallel)$ is much less than unity at most wavenumbers in Zone (1), whereas $\chi_{B_A}(k_\perp,k_\parallel)$ increases towards unity and follows the scaling $k_\parallel\propto k_\perp^{2/3}$ in Zone (3). These results suggest a transition from weak to strong nonlinear interactions, agreeing with theoretical expectations and simulations \cite{Howes2011,Verdini2012,Meyrand2016}. 

With the measurements of proton velocity fluctuations, we observe a similar Alfvénic weak-to-strong transition (Supplementary Fig. 7). The transition scale ($\lambda_{CB}$) is estimated by the smallest perpendicular wavenumber of strong turbulence ($k_{\perp,CB}$), where $\lambda_{CB}\approx 1/k_{\perp,CB}$. For both magnetic field and proton velocity fluctuations, $k_{\perp,CB}$ is around $3\times10^{-4}km^{-1}$, marked by the second vertical lines in Fig. 3 and Supplementary Fig. 7. The consistency in the transition scales estimated by magnetic field and proton velocity measurements further confirms the reliability of our findings.

A notable perturbation is present in Zone (2), as a result of local enhancements of magnetic energy at $k_\perp\approx 1.8\times10^{-4}km^{-1}$ (Fig. 2), leading to the simultaneous existence of strong nonlinearity ($\chi_{B_A}\approx 1$) and weak nonlinearity ($\chi_{B_A}\ll 1$) in the wave number range corresponding to those in Fig.3b. Thus, the Alfvénic weak-to-strong transition more likely occurs within a ‘region’ rather than at a critical wavenumber. Besides, we do not discuss the fluctuations in Zone (4). The deviations of data sets under $\eta<10^\circ$ and $\eta<15^\circ$ in Zone (3) of Fig. 3b are likely due to the limited data samples (Supplementary Fig. 6). The uncertainties mentioned above do not affect our main conclusions.

Fig. 4 presents $k_\perp$ versus $f_{rest}$ distributions of magnetic energy, where $f_{rest}$ is the frequency in the plasma flow frame. At $k_\perp<5\times10^{-5}km^{-1}$, magnetic energy is concentrated at $f_{rest}\approx f_A$, where $f_A$ is Alfvén frequency (horizontal dotted lines with error bars). At $k_\perp>1\times10^{-4}km^{-1}$, the range of $f_{rest}$ broadens, mostly deviating from $f_A$. Nevertheless, the boundary of fluctuating frequencies is roughly consistent with the scaling $f_{rest}\propto k_\perp^{2/3}$ (the dashed line), indicating that magnetic energy at these wavenumbers satisfies the scaling $k_{\parallel}\propto k_\perp^{2/3}$ due to $f_{rest}\propto k_\parallel$ for Alfvén modes. These results suggest that Alfvénic fluctuations with strong nonlinear interactions do not agree with linear dispersion relations but satisfy the wavenumber scaling of Alfvén modes. The change from single-frequency to broadening-frequency fluctuations with increasing $k_\perp$ suggests a possible transition of turbulence regimes. 

\section{Discussion}

The Alfvénic transition of weak to strong turbulence during cascades to smaller scales is one of the cornerstones of the modern MHD theory. Despite being proposed decades ago, evidence for confirming the existence of the Alfvénic transition is lacking. In this paper, we present direct evidence of the Alfvénic transition via different angles: e.g., the transition of energy spectra (Fig. 3a), Goldreich-Sridhar type envelope for the nonlinear parameter (Fig. 3c), and the spread of $f_{rest}$ on small scales (Fig. 4; See Tab. 1 for a summary). Our observation demonstrates that the Alfvénic transition to strong turbulence is bound to occur with the increase of nonlinearity even fluctuations on large scales are considered as "small amplitude" ($M_{A,turb}\approx 0.33$). We want to point out that plasma parameters in the analyzed event are generic, and Alfvénic weak-to-strong transition can occur in other astrophysical and space plasma systems. The impact of our findings goes beyond the study of turbulence itself to particle transport and acceleration \cite{Schlickeiser02,Yan2022rev}, magnetic reconnection \cite{Matthaeus1986,LV99}, star formation \cite{Crutcher2012,Padoan2014}, and all the other relevant fields \cite[see, e.g.][]{Zhang2011, HY09}. 

\vspace{0.5cm}
\noindent\textbf{Method}

\noindent\textbf{Geocentric-solar-ecliptic (GSE) coordinates}

\noindent We use the GSE coordinates in this study. $X_{GSE}$ points towards the Sun from the Earth, $Z_{GSE}$ orients along the ecliptic north pole, and $Y_{GSE}$ completes a right-handed system.

\noindent\textbf{Trace power spectral densities}

\noindent The trace power spectral densities of magnetic field and proton velocity ($P_B=P_{B,X}+P_{B,Y}+P_{B,Z}$ and $P_V=P_{V,X}+P_{V,Y}+P_{V,Z}$) are calculated by applying the fast Fourier transform with three-point centered smoothing in GSE coordinates. We choose the intermediate instant of each time window as the time point where the spectral slope varies with time. 

\noindent\textbf{Alfvén mode decomposition method}

\noindent We calculate wavenumber-frequency distributions of Alfvénic magnetic field and proton velocity power by an improved Alfvén mode decomposition method. This method combines the linear decomposition method \cite{Cho2003}, singular value decomposition (SVD) method \cite{Santolik2003}, and multi-spacecraft timing analysis \cite{Pincon1998}. We perform the calculations in each moving time window with a five-hour length and five-minute moving step. The window length selection (5 hours) provides low-frequency (large-scale) measurements while ensuring $\mathbf{B}_0$ is approaching the local background magnetic field.

\noindent First, we obtain wavelet coefficients ($W$) of magnetic field and proton velocity using Morlet-wavelet transforms \cite{Grinsted2004}. To eliminate the edge effect due to finite-length time series, we perform wavelet transforms twice the time window length and cut off the affected periods.

\noindent Second, wavevector directions $(\mathbf{k}_{SVD}(t,f_{sc}))$ are determined by SVD of magnetic wavelet coefficients \cite{Santolik2003}. The SVD method creates a real matrix equation ($\mathbf{S}\cdot \hat{\mathbf{k}}_{SVD}=0$) equivalent to the linearized Gauss’s law for magnetism ($\mathbf{B} \cdot \hat{\mathbf{k}}_{SVD}=0$). Notice that the minimum singular value of the real matrix $\mathbf{S}$ ($6\times3$) is the best estimate of wavevector directions but cannot determine the wavenumbers. Since relative satellite separations are much shorter than the half-wavelength of MHD scales, the properties of fluctuations simultaneously measured by four \textit{Cluster} spacecraft are similar. Thus, the average wavevector direction and background magnetic field are given by $\mathbf{k}_{SVD}=\frac{1}{4}\sum_{i=1,2,3,4}\hat{\mathbf{k}}_{SVD,Ci}$ and $\mathbf{B}_0=\frac{1}{4}\sum_{i=1,2,3,4}\mathbf{B}_{0,Ci}$. $Ci$ denotes the four \textit{Cluster} spacecraft. 

\noindent Third, we extract Alfvénic components from proton velocity fluctuations based on their incompressibility ($\hat{\mathbf{k}}_{SVD}\cdot\delta\mathbf{V}_p=0$) and perpendicular fluctuating directions ($\hat{\mathbf{b}}_0\cdot\delta\mathbf{V}_p=0$) in wavevector space, where $\delta \mathbf{V}_p$ is expressed by vectors of velocity wavelet coefficients, $\hat{\mathbf{k}}_{SVD}=\mathbf{k}_{SVD}/|\mathbf{k}_{SVD}|$, and $\hat{\mathbf{b}}_0=\mathbf{B}_0/|\mathbf{B}_0|$. Similarly, Alfvénic magnetic field fluctuations are extracted by $\hat{\mathbf{k}}_{SVD}\cdot\delta \mathbf{B}=0$ and $\hat{\mathbf{b}}_0\cdot\delta \mathbf{B}=0$, according to the linearized induction equation
\begin{eqnarray}
\omega\delta\mathbf{B}=\mathbf{k}\times(\mathbf{B}_0\times\delta\mathbf{V}_p)\approx|\mathbf{k}|\hat{\mathbf{k}}_{SVD}\times(\mathbf{B}_0\times\delta\mathbf{V}_p)
\label{4},
\end{eqnarray}
where $\mathbf{k}$ is the wavevector. Thus, Alfvénic proton velocity and magnetic field fluctuations are in the same direction $\hat{\mathbf{k}}_{SVD}\times\hat{\mathbf{b}}_0/|\hat{\mathbf{k}}_{SVD}\times\hat{\mathbf{b}}_0|$ (see Schematic in Supplementary Fig. 4).

\noindent Fourth, Alfvénic magnetic power at each time $t$ and $f_{sc}$ is calculated by $P_{B_A}(t,f_{sc})=\frac{1}{4}\sum_{i=1,2,3,4}W_{B_A,Ci}W_{B_A,Ci}^*$. Alfvénic proton velocity power is calculated by $P_{V_A}(t,f_{sc})=W_{V_A,C1}W_{V_A,C1}^*$. This is because magnetic field data are available on four \textit{Cluster} spacecraft, whereas proton plasma data are only available on \textit{Cluster}-1 during the analyzed period.

\noindent Fifth, noticing that SVD does not give the magnitude of wavevectors, we calculate wavevectors $(\mathbf{k}_A(t,f_{sc}))$ using the multispacecraft timing analysis based on phase differences between the Alfvénic magnetic field from four spacecraft \cite{Pincon1998}. Magnetic field data are interpolated to a uniform time resolution of $8 samples/s$ for sufficient time resolutions. We consider that the wave front is moving in the direction $\mathbf{\hat{n}}$ with velocity $V_{w}$. The wavevectors $\mathbf{k}_A=2\pi f_{sc}\mathbf{m}$, where the vector $\mathbf{m}=\mathbf{\hat{n}}/V_{w}$, and the subscript $A$ represent the Alfvénic component.  
\begin{eqnarray}
  \left(
  \begin{array}{cc}
      \mathbf{r}_{2} - \mathbf{r}_{1} \\
      \mathbf{r}_{3} - \mathbf{r}_{1} \\
      \mathbf{r}_{4} - \mathbf{r}_{1}
  \end{array} 
  \right) \mathbf{m} = 
  \left(
 \begin{array}{cc}
      \delta t_2 \\
      \delta t_3 \\
      \delta t_4 
 \end{array} 
  \right)
  \label{eq:1}
\end{eqnarray}
where \textit{Cluster}-1 has arbitrarily been taken as the reference. The left side of Eq.(\ref{eq:1}) is the relative spacecraft separations. The right side of Eq.(\ref{eq:1}) represents the weighted average time delays, estimated by the ratio of six phase differences ($\phi_{ij}=arctan(\mathcal{S}(W_{B_A}^{ij}),\mathcal{R}(W_{B_A}^{ij}))$) to the angular frequencies ($\omega_{sc}=2\pi f_{sc}$), where $\phi_{ij}$ is from all spacecraft pairs ($ij=12$, $13$, $14$, $23$, $24$, $34$)). $\mathcal{S}$ and $\mathcal{R}$ are the imaginary and real parts of cross-correlation coefficients, respectively. Four \textit{Cluster} spacecraft provide six cross-correlation coefficients \cite{Grinsted2004}, i.e., $W_{B_A}^{12}=\langle W_{B_A,C1}W_{B_A,C2}^*\rangle$, $W_{B_A}^{13}=\langle W_{B_A,C1}W_{B_A,C3}^*\rangle$, $W_{B_A}^{14}=\langle W_{B_A,C1}W_{B_A,C4}^*\rangle$, $W_{B_A}^{23}=\langle W_{B_A,C2}W_{B_A,C3}^*\rangle$, $W_{B_A}^{24}=\langle W_{B_A,C2}W_{B_A,C4}^*\rangle$, and $W_{B_A}^{34}=\langle W_{B_A,C3}W_{B_A,C4}^* \rangle$, where $\langle...\rangle$ denotes a time average over $256s$ for the reliability of phase differences.

It is worth noting that timing analysis determines the actual wavevectors of the Alfvénic magnetic field. In contrast, the SVD method determines the best estimate of the wavevector sum in three magnetic field components \cite{Santolik2003}. Thus, $\mathbf{k}_A$ is not completely aligned with $\hat{\mathbf{k}}_{SVD}$. Besides, we restrict our analysis to fluctuations with small angle $\eta$ between $\hat{\mathbf{k}}_{SVD}$ and $\mathbf{k}_A$, to ensure the reliability of the extraction process (the third step). With relaxed $\eta$ constraints, more sampling points are involved; thus, the uncertainty from limited measurements decreases. On the other hand, with relaxed $\eta$ constraints, $k_{A}$ deviates more from $k_{SVD}$, which may increase the uncertainty. This letter presents results from five data sets under $\eta<10^\circ$, $\eta<15^\circ$, $\eta<20^\circ$, $\eta<25^\circ$, and $\eta<30^\circ$ to investigate the effects of uncertainties introduced by the combination of the SVD method and timing analysis.

\noindent Sixth, we construct a set of $400\times 400\times 400$ bins to obtain wavenumber-frequency distributions of magnetic power ($P_{B_A}(k_\perp, k_\parallel,f_{sc})$) and proton velocity power ($P_{V_A}(k_\perp,k_\parallel,f_{sc})$), where the parallel wavenumber is $k_\parallel=\mathbf{k}_A\cdot\hat{\mathbf{b}}_0$, and the perpendicular wavenumber is $k_\perp=\sqrt{\mathbf{k}_A^2-k_\parallel^2}$. Each bin subtends approximately the same $k_\perp$, $k_\parallel$, and $f_{sc}$. To cover all MHD wavenumbers and ensure measurement reliability, we restrict our analysis to fluctuations with $1/(100d_{sc})<k<min(0.1/max(d_i,r_{ci}),\pi/d_{sc})$ and $2/t^*<f_{rest}<f_{ci}/2$, and fluctuations beyond these wavenumber and frequency ranges are set to zero. Here, $d_{sc}$ is relative satellite separations, $min(*)$ and $max(*)$ are the minimum and maximum, $d_i$ is the proton inertial length, $r_{ci}$ is the proton gyro-radius, $t^*$ is the duration studied, $f_{rest}=f_{sc}-\mathbf{k}_A\cdot \mathbf{V}_p/(2\pi)$ is the frequency in the plasma flow frame, and $\mathbf{V}_p$ is the proton bulk velocity with the spacecraft velocity being negligible. This study utilizes the representation of absolute frequencies:
\begin{equation}
   (f_{rest},\mathbf{k}_A)=
   \begin{cases}
        (f_{rest},\mathbf{k}_A) &  f_{rest}>0  \\
        (-f_{rest},-\mathbf{k}_A) &  f_{rest}<0 
   \end{cases}
\end{equation}

$P_{\epsilon_A}(k_\perp, k_\parallel,f_{sc})$ are obtained by averaging $P_{\epsilon_A}(k_\perp,k_\parallel,f_{sc},t)$ over effective time points in all time windows at each $f_{sc}$ and each $\mathbf{k}$, where $\epsilon=V,B$ represents the proton velocity ($V$) and magnetic field ($B$). 

\noindent\textbf{Alfvén speed units}

\noindent For comparison, this study presents the fluctuating magnetic field in Alfvén speed units, which is normalized by $\sqrt{\mu_0m_pN_0}$, where $\mu_0$ is the vacuum permeability, $m_p$ is the proton mass, and $N_0$ is the mean proton density.

\noindent\textbf{magnetic energy spectral density}

\noindent This study defines the energy spectral density of magnetic field as $E_{B_A}(k_\perp)=\frac{1}{2}\frac{\delta B_A^2(k_\perp)}{k_\perp}$, where the Alfvénic magnetic energy density is calculated by $\delta B_A^2(k_\perp)=2\sum_{k_\perp=k_\perp}^{k_\perp\rightarrow\infty}\sum_{k_\parallel=0}^{k_\parallel\rightarrow\infty}\int_0^\infty P_{B_A}(k_\perp,k_\parallel,f_{sc})df_{sc}$.

\noindent\textbf{Nonlinearity parameter}

The nonlinearity parameter is estimated by $\chi_{B_A} (k_\perp,k_\parallel) \approx k_\perp \delta B_A(k_\perp,k_\parallel)/(k_\parallel B_0)$, where the Alfvénic magnetic energy density is calculated by $\delta B_A^2(k_\perp,k_\parallel)=\sum_{k_\perp=k_\perp}^{k_\perp\rightarrow\infty}\sum_{k_\parallel=k_\parallel}^{k_\parallel\rightarrow\infty}\int_0^\infty P_{B_A}(k_\perp,k_\parallel,f_{sc})df_{sc}$, and $B_0$ in Alfvén speed units is around $106km/s$. 

\noindent\textbf{Frequency-wavenumber distribution of magnetic energy}

\noindent The frequency-wavenumber distributions of magnetic energy is approximately estimated by $D_{B_A}(k_\perp,f_{sc})\approx\sum_{k_\parallel=0}^{k_\parallel\rightarrow\infty}P_{B_A}(k_\perp,k_\parallel,f_{sc})\Delta f_{sc}$ and is transformed into the plasma flow frame by correcting the Doppler shift $f_{rest}=f_{sc}-\mathbf{k}_A\cdot \mathbf{V}/(2\pi)$.

\section*{Data Availability}

\noindent The Cluster data are available at \url{https://cdaweb.gsfc.nasa.gov}.

\section*{Code Availability}

\noindent Data analysis was performed using the IRFU-MATLAB analysis package available at \url{https://github.com/irfu/irfu-matlab}. 

\section*{Acknowledgments}

\noindent We would like to thank the members of the \textit{Cluster} spacecraft team and NASA’s Coordinated Data Analysis Web. K.H.Y. is supported by the Laboratory Directed Research and Development program of Los Alamos National Laboratory grant 20220700PRD1.

\section*{Author contributions}

\noindent H.Y. initiated and designed the project. S.Z. and T.Z.L. designed and completed the data processing methods. S.Z. carried out the specific observation data processing. S.Z., H.Y., T.Z.L., K.H.Y., and H.W. contributed to the theoretical analysis of the main results. All authors contributed to writing, editing, and approving the manuscript.

\section*{Competing interests}

\noindent The authors declare no competing interests.

\begin{table}[h]
\caption{Transition wavenumbers are determined by magnetic field measurements.}\label{tab1}%
\begin{tabular}{@{}lll@{}}
\toprule
      &Weak MHD turbulence & Strong MHD turbulence\\
\midrule
 $k_\parallel -k_\perp$ distributions of & Purely perpendicular cascade & Goldreich-Sridhar cascade \\
magnetic energy & $k_\perp<2\times10^{-4}km^{-1}$ & $k_\perp>2\times10^{-4}km^{-1}$\\ \hline
Spectral slopes of &Wave-like ($-2$) & Kolmogorov-like ($-5/3$) \\
magnetic energy & $1.6\times10^{-4}<k_\perp<3\times10^{-4} km^{-1}$ & $3\times10^{-4}<k_\perp<7\times10^{-4} km^{-1}$ \\ \hline
 Nonlinearity parameter &$\chi_{BA}\ll 1$ & $\chi_{BA}\approx1$ and $\chi_{BA}\geq 1$ \\
 ($\chi_{BA}$) & $k_\perp<1\times10^{-4}km^{-1}$ & $3\times10^{-4}<k_\perp< 7\times10^{-4}km^{-1}$ \\ \hline
 Frequency-wavenumber &  Single-frequency fluctuations &Broadening-frequency fluctuations \\ 
distributions & $f_{rest}\approx f_A$ &with $f_{rest}\propto k_\perp^{2/3}$ boundary \\
& $k_\perp<5\times10^{-5}km^{-1}$ & $k_\perp>1\times10^{-4}km^{-1}$\\ \hline
\botrule
\end{tabular}
\end{table}

\newcounter{Sfigure}
\setcounter{Sfigure}{1}
\renewcommand{\thefigure}{Fig.\arabic{Sfigure}}
\begin{figure}[h]
\centering
\includegraphics[width=0.8\textwidth]{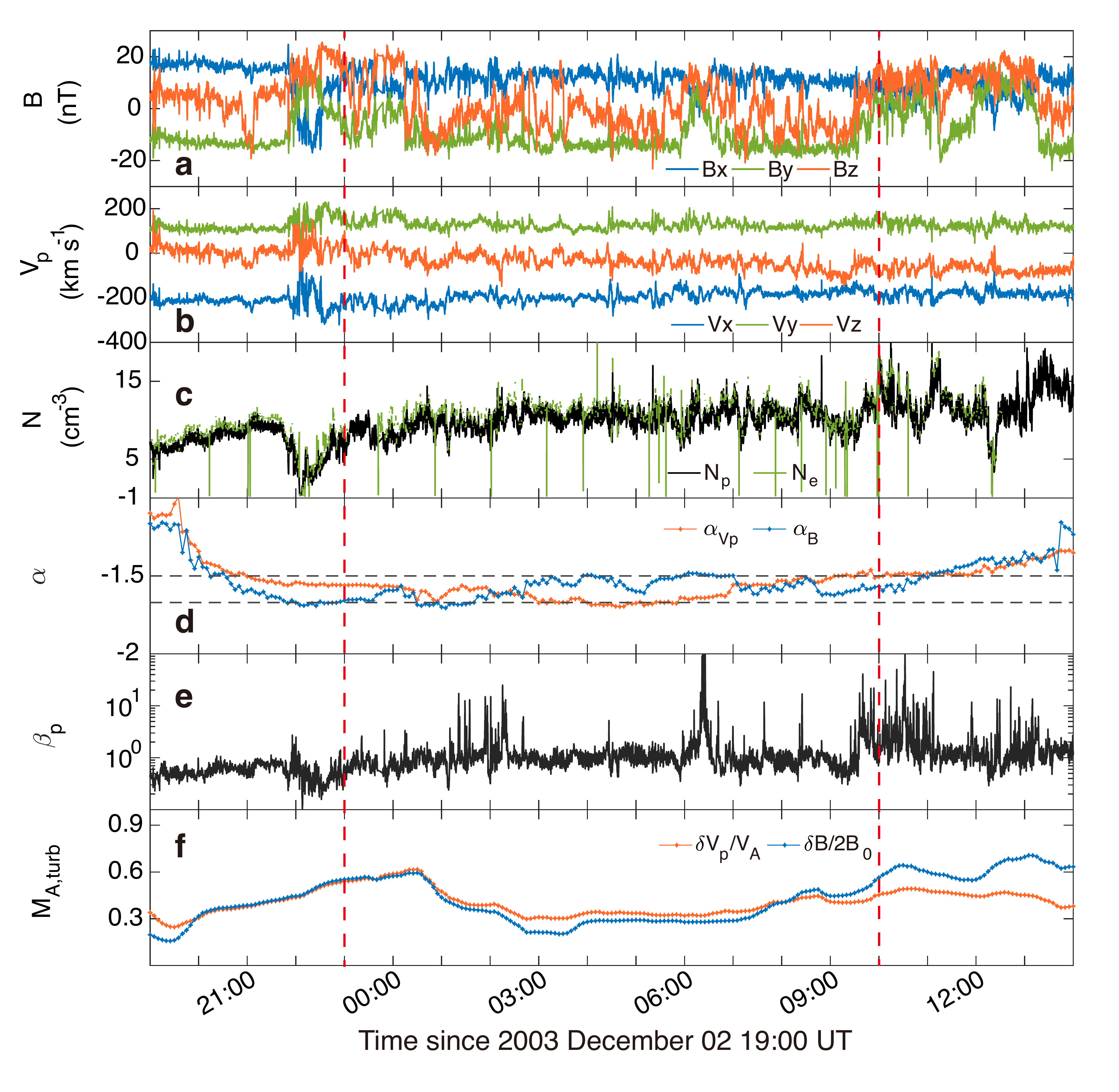}
\caption{An overview of fluctuations measured by \textit{Cluster}-1 in Earth's magnetosheath on 2-3 December 2003. The data are displayed in GSE coordinates. \textbf{a}, Magnetic field components ($B_X$, $B_Y$ and $B_Z$). \textbf{b}, Proton bulk velocity ($V_X$, $V_Y$ and $V_Z$). \textbf{c}, Proton and electron density. \textbf{d}, Spectral slopes ($\alpha$) of magnetic field and proton velocity fluctuations between $0.001 Hz$ and $0.1f_{ci}$. The two horizontal lines represent $\alpha=-5/3$ and $-3/2$. \textbf{e}, The proton plasma $\beta_p$. \textbf{f}, The turbulent Alfvén Mach number ($M_{A,turb}=\delta V_p/V_A$) and half of the relative amplitudes of the magnetic field ($\delta B/(2B_0)$), where $\delta V_p$ and $\delta B$ are rms proton velocity and magnetic field fluctuations, respectively. The fluctuations analyzed in detail are during 23:00-10:00 UT on 2-3 December, marked between the two vertical dashed lines.}\label{Fig1}
\end{figure}

\setcounter{Sfigure}{2}
\renewcommand{\thefigure}{Fig.\arabic{Sfigure}}
\begin{figure}[h]
\centering
\includegraphics[width=1\textwidth]{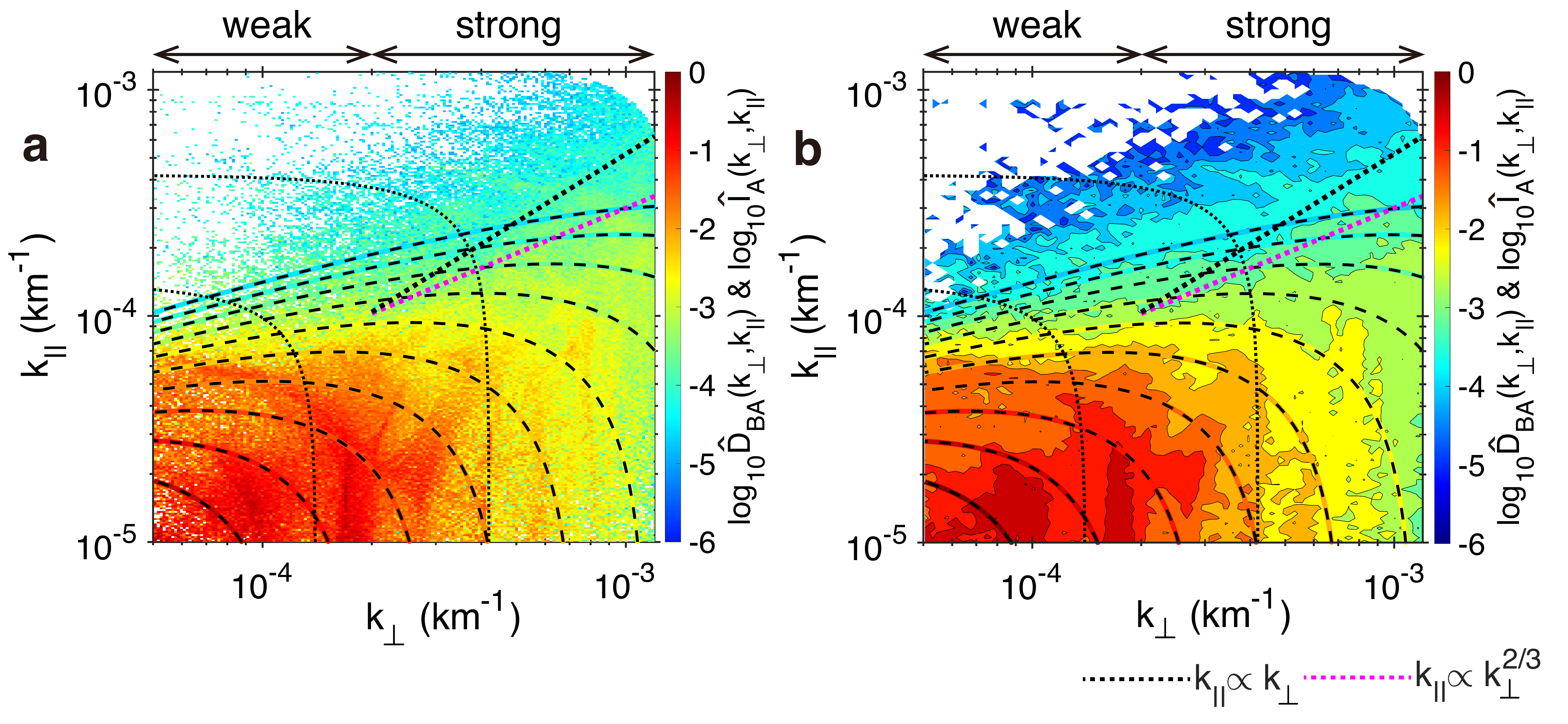}
\caption{The comparison between wavenumber distributions of Alfvénic magnetic energy ($\hat{D}_{B_A}(k_\perp,k_\parallel)$) and theoretical energy spectra ($\hat{I}_A(k_\perp,k_\parallel)$). \textbf{a}, 2D spectral image of $\hat{D}_{B_A}(k_\perp,k_\parallel)$ with a high resolution ($400\times400$ bins). \textbf{b}, 2D filled contours of $\hat{D}_{B_A}(k_\perp,k_\parallel)$ with low-resolution binning ($150\times150$ bins), to clarify the contours. \textbf{a},\textbf{b}, $\hat{I}_A(k_\perp,k_\parallel)$ at $L_0\approx4.6\times10^4km$ is displayed by color contours with black dashed curves, which is in the same color map as $\hat{D}_{B_A}(k_\perp,k_\parallel)$. The black dotted curves mark $k=\sqrt{k_\parallel^2+k_\perp^2}=0.01/d_i$ and $0.03/d_i$. These figures utilize the data set under $\eta<30^\circ$.}\label{Fig2}
\end{figure}

\setcounter{Sfigure}{3}
\renewcommand{\thefigure}{Fig.\arabic{Sfigure}}
\begin{figure}[h]
\centering
\includegraphics[width=1\textwidth]{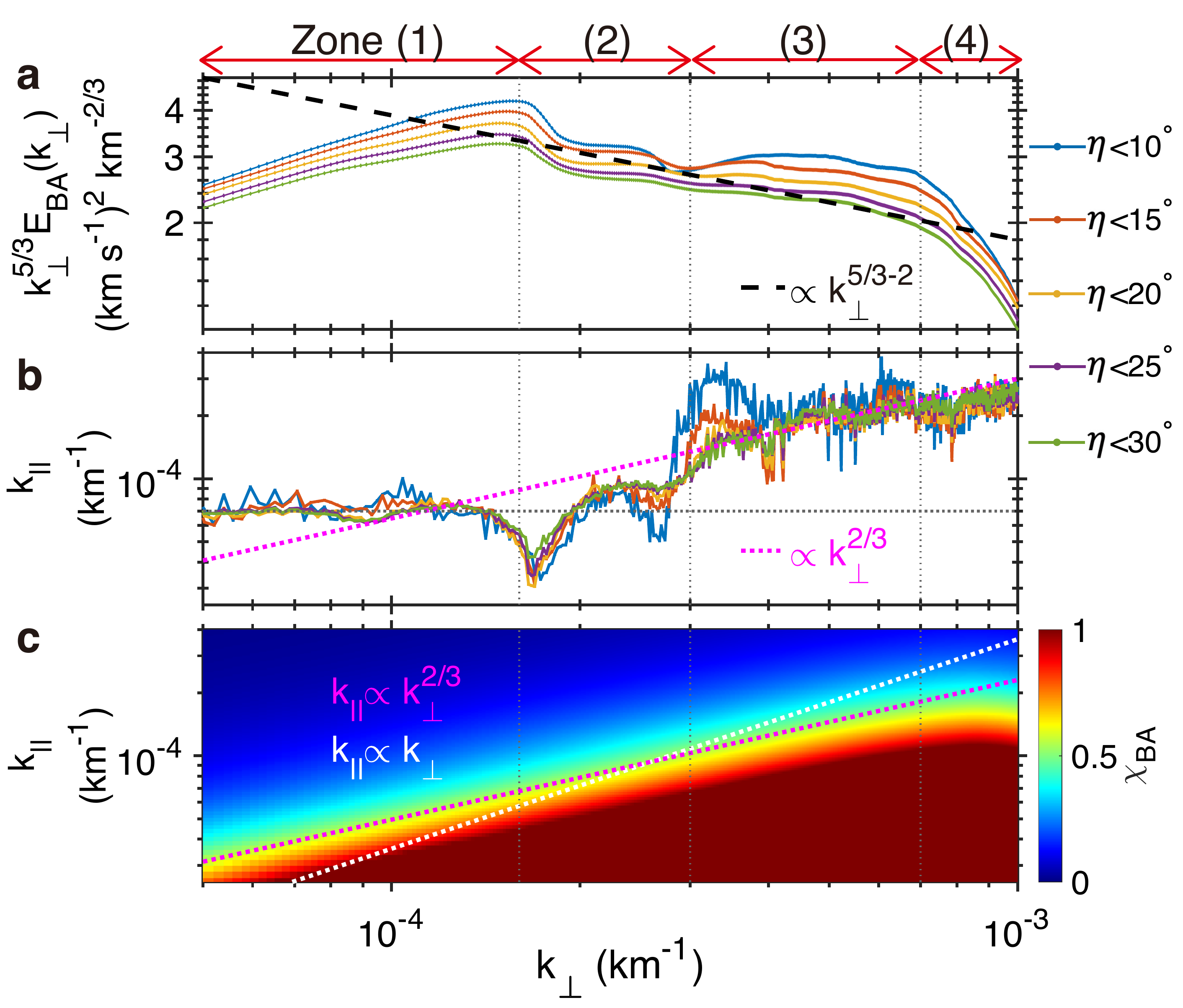}
\caption{Perpendicular wavenumber dependence of the compensated spectra ($k_\perp^{5/3}E_{B_A}(k_\perp)$), parallel wavenumber ($k_\parallel$), and nonlinearity parameter ($\chi_{B_A}(k_\perp,k_\parallel)$). \textbf{a}, $k_\perp^{5/3}E_{B_A}(k_\perp)$ are displayed by curves. The dashed line represents the scaling $k_\perp^{5/3}E_{B_A}(k_\perp)\propto k_\perp^{5/3-2}$. To facilitate comparison with proton velocity fluctuations, magnetic field fluctuations are in Alfvén speed units. \textbf{b}, The variation of $k_\parallel$ versus $k_\perp$. The dashed line represents the scaling $k_\parallel \propto k_\perp^{2/3}$. The horizontal dotted line marks $k_\parallel=7\times10^{-5} km^{-1}$. \textbf{c}, $\chi_{B_A}(k_\perp,k_\parallel)$ spectrum calculated using the data set under $\eta<30^\circ$. The $k_{\perp}$ dependence figure is divided into four zones: (1) $5\times10^{-5}<k_\perp<1.6\times10^{-4}km^{-1}$, (2) $1.6\times10^{-4}<k_\perp<3\times10^{-4}km^{-1}$, (3) $3\times10^{-4}<k_\perp<7\times10^{-4}km^{-1}$, and (4) $7\times10^{-4}<k_\perp<1\times10^{-3}km^{-1}$. The first, second, and third vertical dotted lines are around the maximum of $k_\perp^{5/3}E_{B_A}(k_\perp)$, the beginning and the end of flattened $k_\perp^{5/3}E_{B_A}(k_\perp)$, respectively.}\label{Fig3}
\end{figure}

\setcounter{Sfigure}{4}
\renewcommand{\thefigure}{Fig.\arabic{Sfigure}}
\begin{figure}[h]
\centering
\includegraphics[width=0.8\textwidth]{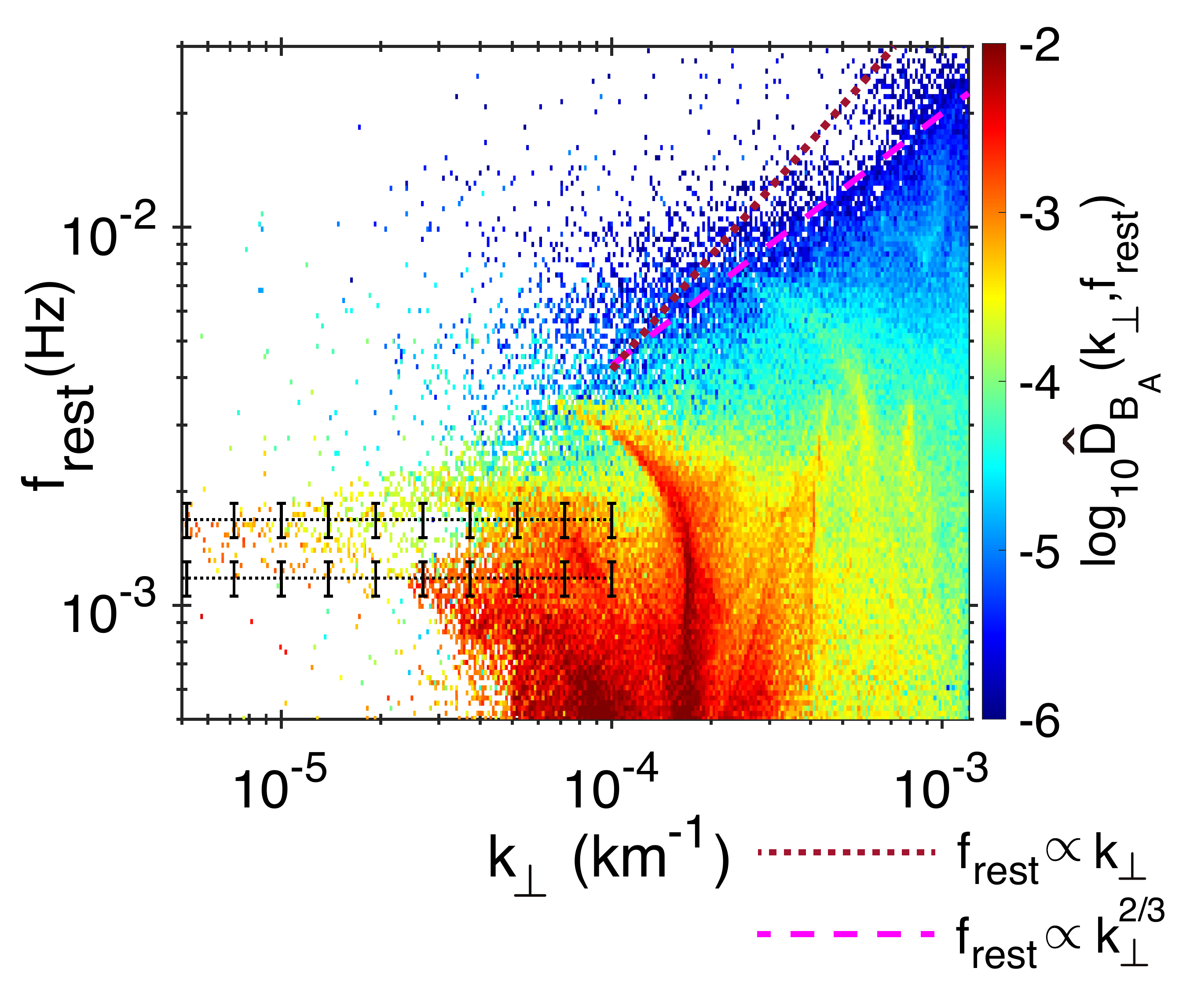}
\caption{The $k_\perp-f_{rest}$ distributions of Alfvénic magnetic energy in the plasma flow frame. $\hat{D}_{B_A}(k_\perp,f_{rest})=D_{B_A}(k_\perp,f_{rest})/D_{B_A,max}$ is normalized by the maximum magnetic energy in all $(k_\perp,f_{rest})$ bins. The horizontal dotted lines represent theoretical Alfvén frequencies $f_A=|k_\parallel V_A|/(2\pi)$, where $k_\parallel\approx7\times10^{-5}km^{-1}$ in Zone (1) of Fig. 3b, and $k_\parallel\approx1\times10^{-4}km^{-1}$ in Zone (1) of Supplementary Fig. 7b. The $f_A$ uncertainties are estimated by the standard deviation of $V_A$ ($106\pm 11 km^{-1}$), illustrated by error bars on corresponding horizontal dotted lines. This figure utilizes the data set under $\eta<30^\circ$.}\label{Fig4}
\end{figure}

\appendix

\section{Deviations between the mean magnetic field and local field}

The anisotropy of Alfvénic fluctuations depends on the local background magnetic field. Although it would be better to use a scale-dependent mean magnetic field ideally, the mode decomposition method is based on a perturbative treatment of fluctuations in the presence of a uniform magnetic field ($\mathbf{B}_0$). This method requires $\mathbf{B}_0$ independent of the transformation between real and wavevector space. Nevertheless, Supplementary Fig. 1 shows the spacecraft-frame frequency-time spectrum of the cosine of angle ($|cos\langle\mathbf{B}_0,\mathbf{B}_{local}\rangle|$) between $\mathbf{B_0}$ and $\mathbf{B}_{local}$. The local mean field is calculated as $\mathbf{B}_{local}=[\mathbf{B}(t-2\tau)+4\mathbf{B}(t-\tau)+6\mathbf{B}(t)+4\mathbf{B}(t+\tau)+\mathbf{B}(t+2\tau)]/16$, where $\tau$ is the timescale. The mean magnetic field ($\mathbf{B}_0$) within a five-hour moving time window is closely aligned with $\mathbf{B}_{local}$, suggesting that $\mathbf{B}_0$ approximating the local mean field is acceptable. 

\section{Two-dimensional energy wavenumber distributions at different time window lengths}

To further address this limitation of the mode decomposition method, this study explores the variation of two-dimensional (2D) wavenumber distributions of Alfvénic magnetic energy $D_{B_A}(k_\perp,k_\parallel)$ by adjusting the length of time windows, where

\begin{eqnarray}
D_{B_A}(k_\perp,k_\parallel)=\int_0^\infty P_{B_A}(k_\perp,k_\parallel,f_{sc})df_{sc}.
\end{eqnarray}

We show magnetic energy spectra with different time window lengths in Supplementary Fig. 2. To simplify, the modeled theoretical energy spectra ($I_A(k_\perp,k_\parallel)$) are estimated with the same parameters ($M_{A,turb}\approx0.33$, $L_0\approx 4.6\times 10^4km$). (i) The longer time window length provides more low-frequency (large-scale) measurements. (ii) Energy spectra with shorter time window lengths are more consistent with theoretical contours in the strong turbulence regime (at larger $k_\perp$). It is likely because the mean magnetic field in shorter time windows is closer to the local mean field of fluctuations with larger wavenumbers. (iii) The main changes in energy distributions are little affected by time window length: $k_\parallel$ distributions of magnetic energy start to broaden around $k_\perp>2\times 10^{-4} km^{-1}$ for all panels. 

We show the results with a five-hour length in the main text for two reasons: (i) The length cannot be too long. The shorter the window length, the closer to the local background magnetic field. (ii) The length cannot be too short in order to ensure the measurements of low-frequency (large-scale) signals since the Alfvénic weak-to-strong transition is present on relatively large scales. The five-hour length selection provides the low-frequency (large-scale) measurements while ensuring $\mathbf{B}_0$ is approaching the local background magnetic field.

\section{Details of examination of the turbulence state}

To examine the turbulent state, we calculate the normalized correlation function $R(\tau)/R(0)$, where the correlation function is defined as $R(\tau)=\langle\delta B(t)\delta B(t+\tau)\rangle$, $\tau$ is the timescale, and angular brackets are a time average over the time window length (5 hours). Supplementary Fig. 3 shows $R(\tau)/R(0)$ for magnetic field $\delta B_{\perp 1}$ and $\delta B_{\perp 2}$ components in field-aligned coordinates. Fluctuations $\delta B_{\perp 1}$ are in $(\hat{\mathbf{b}}_0\times\hat{\mathbf{X}}_{GSE})\times\hat{\mathbf{b}}_0$ directions, and $\delta B_{\perp 2}$ are in $\hat{\mathbf{b}}_0\times\hat{\mathbf{X}}_{GSE}$ directions, where $\hat{\mathbf{X}}_{GSE}$ is the unit vector towards the Sun from the Earth.  

This study estimates the correlation time $T_c \approx \int ^{R(\tau)\rightarrow\frac{1}{2e}}_0 R(\tau)/R(0)d\tau$. In Supplementary Fig. 3, $T_c\approx[1300,2300]s$ is much less than the time window length (5 hours), suggesting that fluctuations are approximately stationary. Moreover, $R(\tau)/R(0)$ profiles in all time windows are similar, suggesting that the starting time of the moving time window has a slight influence on $R(\tau)/R(0)$, and thus fluctuations are homogeneous. Above all, it is reasonable to describe structures of turbulent fluctuations using three-dimensional energy distributions.

\section{Schematic of Alfvén mode decomposition from turbulent fluctuations}

Supplementary Fig. 4 shows a coordinate determined by unit vectors of the wavevector and background magnetic field ($\hat{\mathbf{k}}_{SVD}$ and $\hat{\mathbf{b}}_0$). The basis vectors of coordinate axes are in $\hat{\mathbf{b}}_0$, $\hat{\mathbf{k}}_{\perp,out of (k_{SVD},b_0) plane}=\frac{\hat{\mathbf{k}}_{SVD}\times\hat{\mathbf{b}}_0}{|\hat{\mathbf{k}}_{SVD}\times\hat{\mathbf{b}}_0|}$, and $\hat{\mathbf{k}}_{\perp,in (k_{SVD},b_0) plane}=\hat{\mathbf{b}}_0\times\hat{\mathbf{k}}_{\perp,out of (k_{SVD},b_0) plane}$ directions. Alfvénic magnetic field and velocity fluctuations are along $\hat{\mathbf{\xi}}_A$ direction, where $\hat{\mathbf{\xi}}_A=\hat{\mathbf{k}}_{\perp,out of (k_{SVD},b_0) plane}$. The wavevectors ($\mathbf{k}_A$) calculated by multispacecraft timing analysis on Alfvénic magnetic field are not completely alighted with $\hat{\mathbf{k}}_{SVD}$. Thus, we set the angle $\eta$ between $\mathbf{k}_A$ and $\hat{\mathbf{k}}_{SVD}$ as a threshold and only analyze the fluctuations inside the cone.

\section{One-dimensional and two-dimensional wavenumber distributions of Alfvénic energy}

One-dimensional (1D) wavenumber distributions of Alfvénic magnetic energy are calculated by
\begin{eqnarray}
D_{B_A}(k_\perp)=\sum_{k_\parallel=0}^{k_\parallel\rightarrow\infty}\int_0^\infty P_{B_A}(k_\perp,k_\parallel,f_{sc})df_{sc},\\
D_{B_A}(k_\parallel)=\sum_{k_\perp=0}^{k_\perp\rightarrow\infty}\int_0^\infty P_{B_A}(k_\perp,k_\parallel,f_{sc})df_{sc}.
\end{eqnarray}

In Supplementary Fig. 5, 1D wavenumber distributions of Alfvénic magnetic energy from data sets under different $\eta$ limits nearly overlap both for $D_{B_A}(k_\perp)$ and $D_{B_A}(k_\parallel)$, where $\eta$ is the angle between $\mathbf{k}_{SVD}$ and $\mathbf{k}_A$ (Supplementary Fig. 4). Due to the limited data samples, 1D wavenumber distributions from data sets with $\eta<10^\circ$ and $\eta<15^\circ$ show significant deviations from others in Supplementary Fig. 5, and more vacant bins exhibit in 2D wavenumber distributions under smaller $\eta$ in Supplementary Fig. 6. More data samples are involved with the relaxation of $\eta$ limits. On the whole, Alfvénic magnetic energy using data sets under different $\eta$ limits shows similar distributions in Supplementary Figs. 5 and 6.

\section{Energy spectra and nonlinear parameters with velocity measurements}

We observe a similar Alfvénic weak-to-strong transition with the measurements of proton velocity fluctuations. The energy spectral density of Alfvénic velocity is defined as $E_{V_A}(k_\perp)=\frac{\delta V_A^2(k_\perp)}{2k_\perp}$, where the Alfvénic velocity energy density is calculated by $\delta V_A^2(k_\perp)=2\sum_{k_\perp=k_\perp}^{k_\perp\rightarrow\infty}\sum_{k_\parallel=0}^{k_\parallel\rightarrow\infty}\int_0^\infty P_{V_A}(k_\perp,k_\parallel,f_{sc})df_{sc}$. Supplementary Fig. 7a shows the sharp change in spectral slopes of $E_{V_A}(k_\perp)$ from wave-like ($-2$) to Kolmogorov-like ($-5/3$). In Supplementary Fig. 7b, for most of the data points, $k_\parallel$ is approximately stable within $k_\parallel\approx[7\times10^{-5},1\times10^{-4}]km^{-1}$ in Zone (1), whereas the variation of $k_\perp$ versus $k_\parallel$ agrees with the scaling $k_\parallel\propto k_\perp^{2/3}$ in Zone (3). The nonlinearity parameter is estimated as $\chi_{V_A}(k_\perp,k_\parallel)=\frac{k_\perp \delta V_A(k_\perp,k_\parallel)}{k_\parallel V_A}$, where the Alfvénic velocity energy density is estimated by $\delta V_A^2(k_\perp,k_\parallel)=\sum_{k_\perp=k_\perp}^{k_\perp\rightarrow\infty}\sum_{k_\parallel=k_\parallel}^{k_\parallel\rightarrow\infty}\int_0^\infty P_{V_A}(k_\perp,k_\parallel,f_{sc})df_{sc}$. Supplementary Fig. 7c shows that, at the corresponding wavenumbers in Supplementary Fig. 7b, $\chi_{V_A}$ is much less than unity in Zone (1), whereas $\chi_{V_A}$ increases approaching unity and follows the scaling $k_\parallel\propto k_\perp^{2/3}$ in Zone (3). 

\section{Summary of methodology limitation}

\noindent (1) This study is restricted to Alfvénic fluctuations at small amplitude, excluding 2D modes (see Main text).

\noindent (2) This study is limited by ion characteristic scales and satellite relative separations ($1/(100d_{sc})<k<min(0.1/max(d_i,r_{ci}),\pi/d_{sc})$; see Methods). 

\noindent (3) This study examines stationary and homogeneous fluctuations, excluding any involvement in turbulence evolution processes (Supplementary information).

\newcounter{Sfigure1}
\setcounter{Sfigure1}{1}
\renewcommand{\thefigure}{Supplementary Fig.\arabic{Sfigure1}}

\begin{figure}
\centering
\includegraphics[width=1\textwidth]{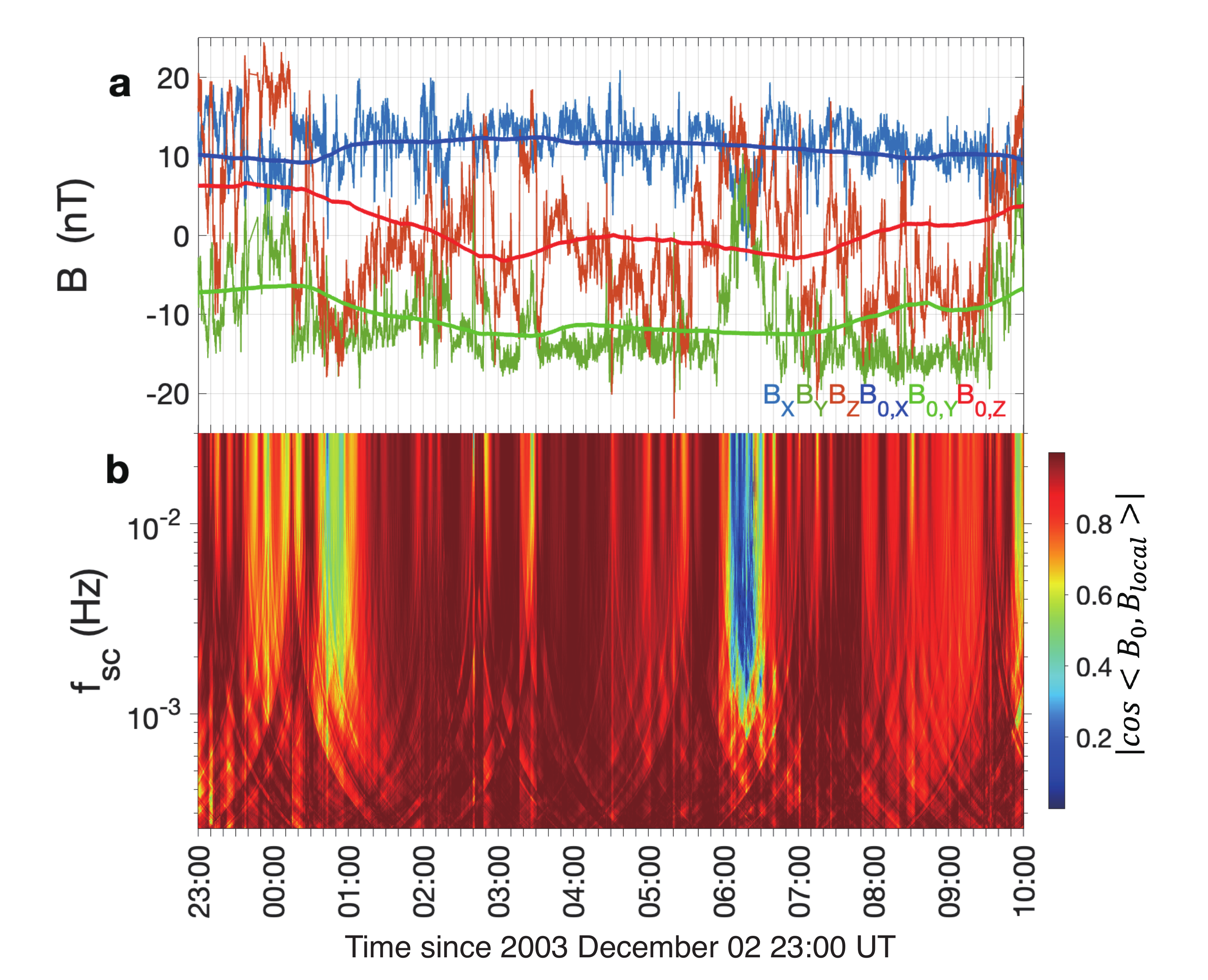}
\caption{Deviations between the mean magnetic field and local field. \textbf{a}, Time series of magnetic field ($\mathbf{B}=[B_X,B_Y,B_Z]$) and mean magnetic field components ($\mathbf{B_0}=[B_{0,X},B_{0,Y},B_{0,Z}]$). \textbf{b}, The spacecraft-frame frequency-time spectrum of the cosine of angle ($|cos\langle\mathbf{B}_0,\mathbf{B}_{local}\rangle|$) between $\mathbf{B_0}$ and $\mathbf{B}_{local}$. $\mathbf{B_0}$ is the mean magnetic field within a five-hour moving time window, and $\mathbf{B}_{local}$ is the local mean magnetic field.}
\end{figure}

\setcounter{Sfigure1}{2}
\renewcommand{\thefigure}{Supplementary Fig.\arabic{Sfigure1}}

\begin{figure}
\centering
\includegraphics[width=1\textwidth]{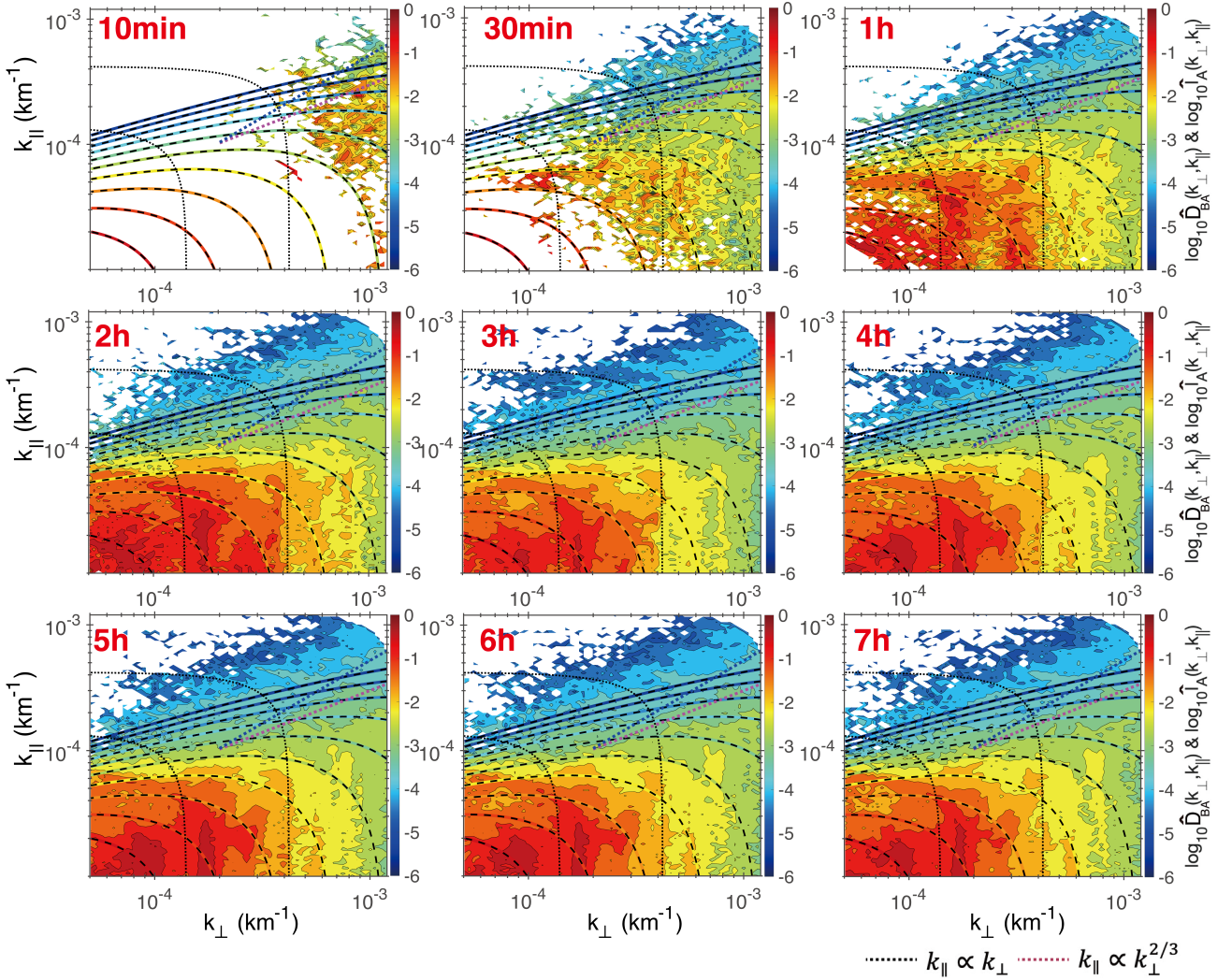}
\caption{The comparisons between wavenumber distributions of Alfvénic magnetic energy $\hat{D}_{B_A}(k_\perp,k_\parallel)$ and theoretical energy spectra $\hat{I}_A(k_\perp,k_\parallel)$. All panels utilize the same format as Fig. 2b in the main text using the data sets under $\eta<30^\circ$. $\hat{D}_{B_A}(k_\perp,k_\parallel)$ is displayed by the filled 2D color contours. $\hat{I}_A(k_\perp,k_\parallel)$ is displayed by color contours with black dashed curves. The time in the upper left corner of each panel represents the length of the time window. The dotted curves mark $k=\sqrt{k_\parallel^2+k_\perp^2}=0.01/d_i$ and $0.03/d_i$.}
\end{figure}

\setcounter{Sfigure1}{3}
\renewcommand{\thefigure}{Supplementary Fig.\arabic{Sfigure1}}

\begin{figure}
\centering
\includegraphics[width=1\textwidth]{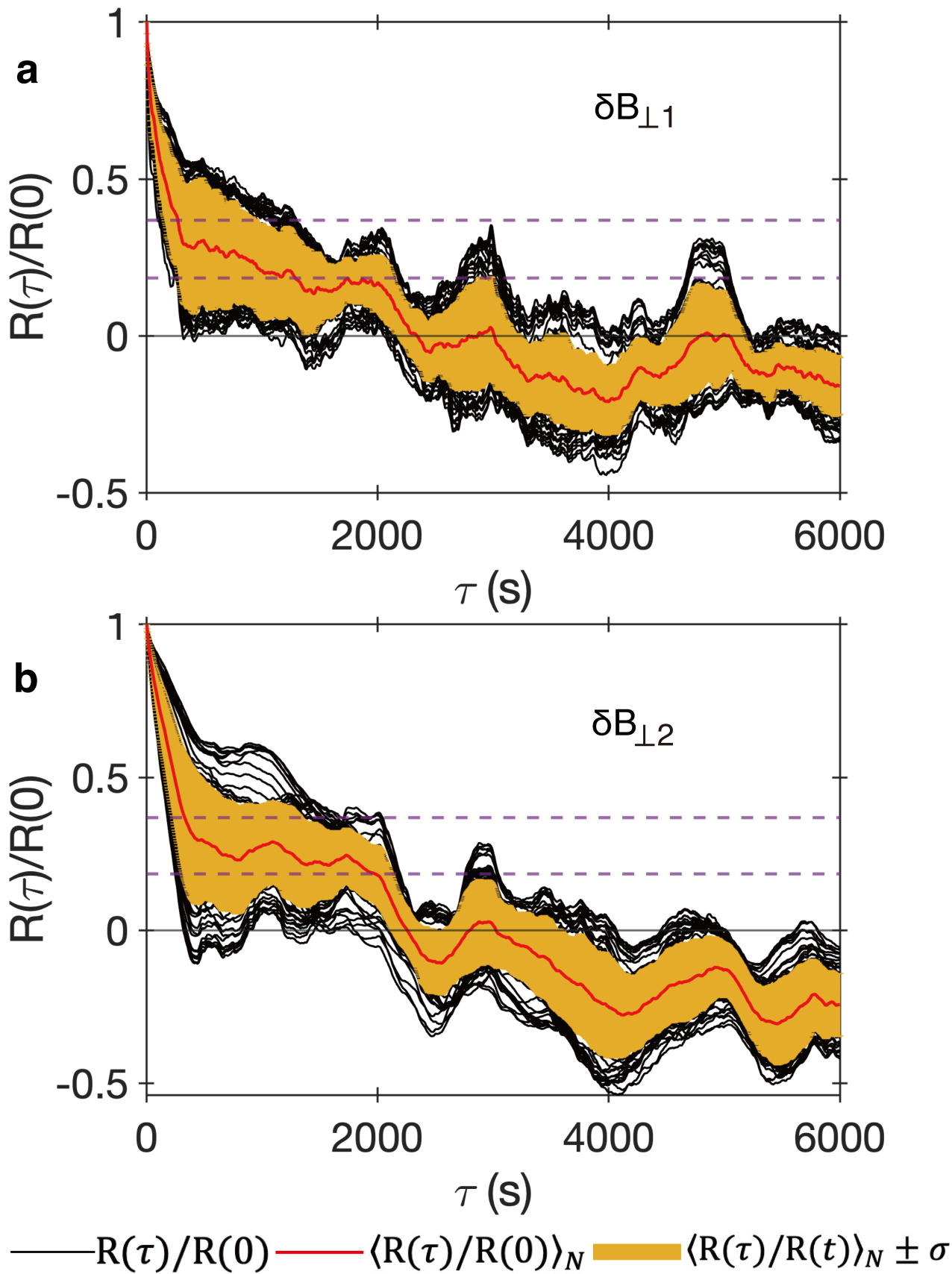}
\caption{Normalized correlation functions ($R(\tau)/R(0)$) versus timescale ($\tau$) in field-aligned coordinates. \textbf{a},\textbf{b}, $R(\tau)/R(0)$ versus $\tau$ for $\delta B_{\perp 1}$ and $\delta B_{\perp 2}$. The solid curves represent $R(\tau)/R(0)$ and average $R(\tau)/R(0)$ ($\langle R(\tau)/R(0)\rangle _N$), and the shaded regions represent $[\langle R(\tau)/R(0)\rangle _N-\sigma,\langle R(\tau)/R(0)\rangle _N+\sigma]$, where $\sigma$ is standard deviations of $R(\tau)/R(0)$ with the window number $N=73$. The horizontal dashed lines represent $R(\tau)/R(0)=1/e$ and $1/(2e)$.}
\end{figure}

\setcounter{Sfigure1}{4}
\renewcommand{\thefigure}{Supplementary Fig.\arabic{Sfigure1}}

\begin{figure}
\centering
\includegraphics[width=1\textwidth]{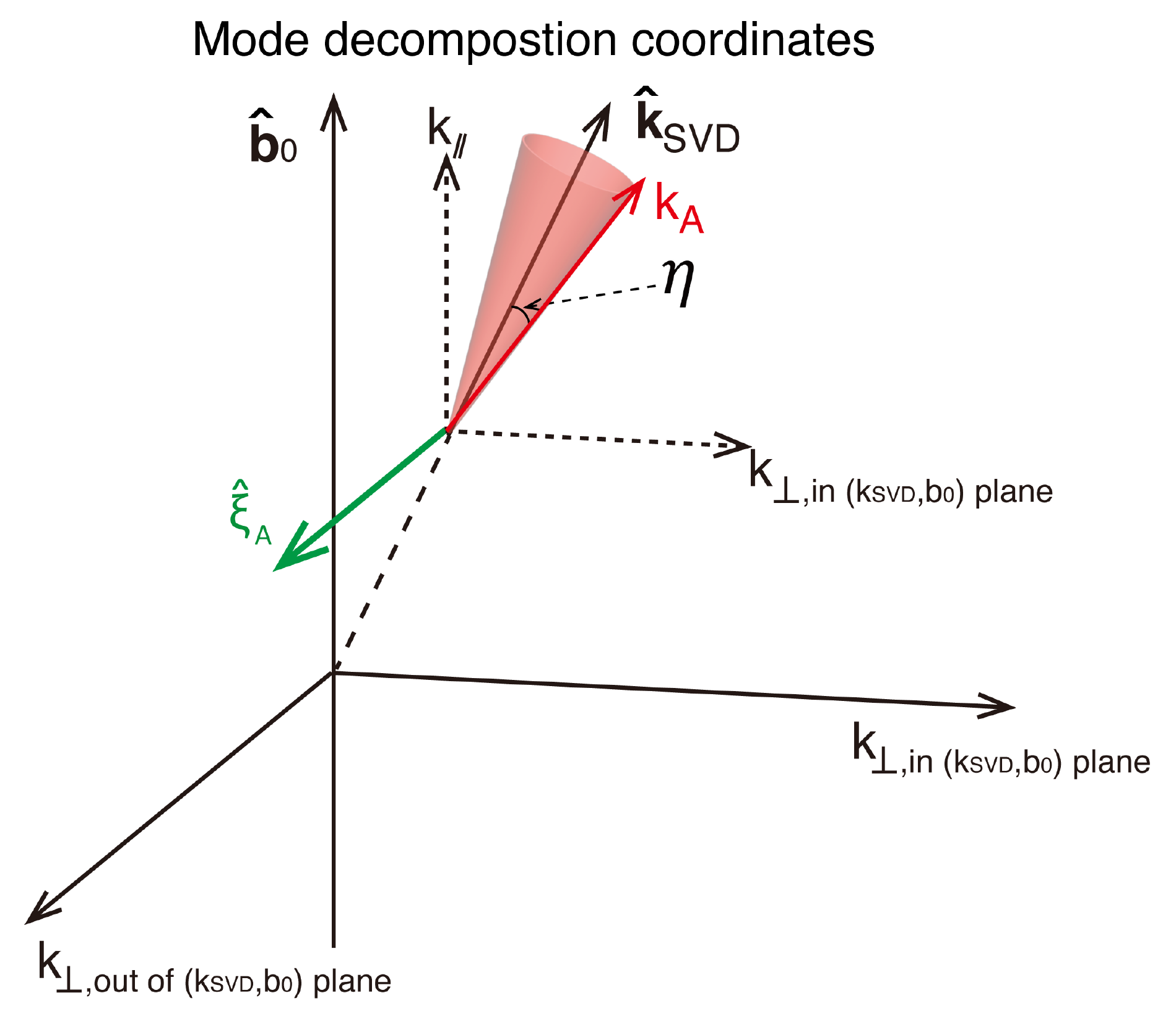}
\caption{Schematic of Alfvén mode decomposition from fluctuations. The coordinates are determined by $\hat{\mathbf{b}}_0$ and $\hat{\mathbf{k}}_{SVD}$. The Alfvénic displacement vector $\hat{\mathbf{\xi}}_A$ is displayed by the arrow along the direction out of $(k_{SVD},b_0)$ plane. The cone marks all $\mathbf{k}_A$ that make an angle $\eta$ with $\hat{\mathbf{k}}_{SVD}$.}
\end{figure}

\setcounter{Sfigure1}{5}
\renewcommand{\thefigure}{Supplementary Fig.\arabic{Sfigure1}}

\begin{figure}
\centering
\includegraphics[width=1\textwidth]{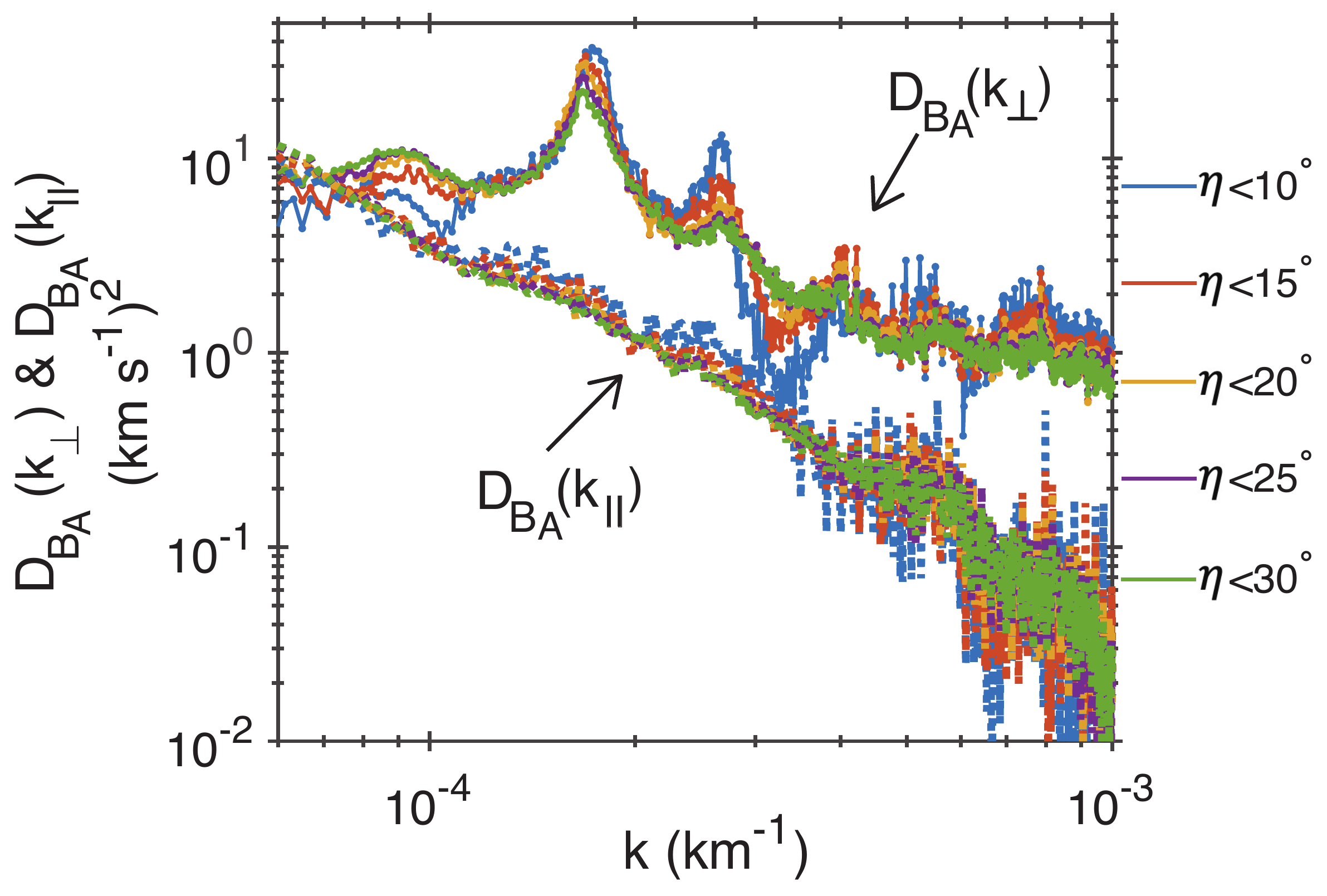}
\caption{1D wavenumber distributions of Alfvénic magnetic energy using data sets under $\eta<10^\circ$, $15^\circ$, $20^\circ$, $25^\circ$, and $30^\circ$. $D_{B_A}(k_\perp)$ is displayed by solid curves. $D_{B_A}(k_\parallel)$ is displayed by dotted curves.}
\end{figure}

\setcounter{Sfigure1}{6}
\renewcommand{\thefigure}{Supplementary Fig.\arabic{Sfigure1}}

\begin{figure}
\centering
\includegraphics[width=1\textwidth]{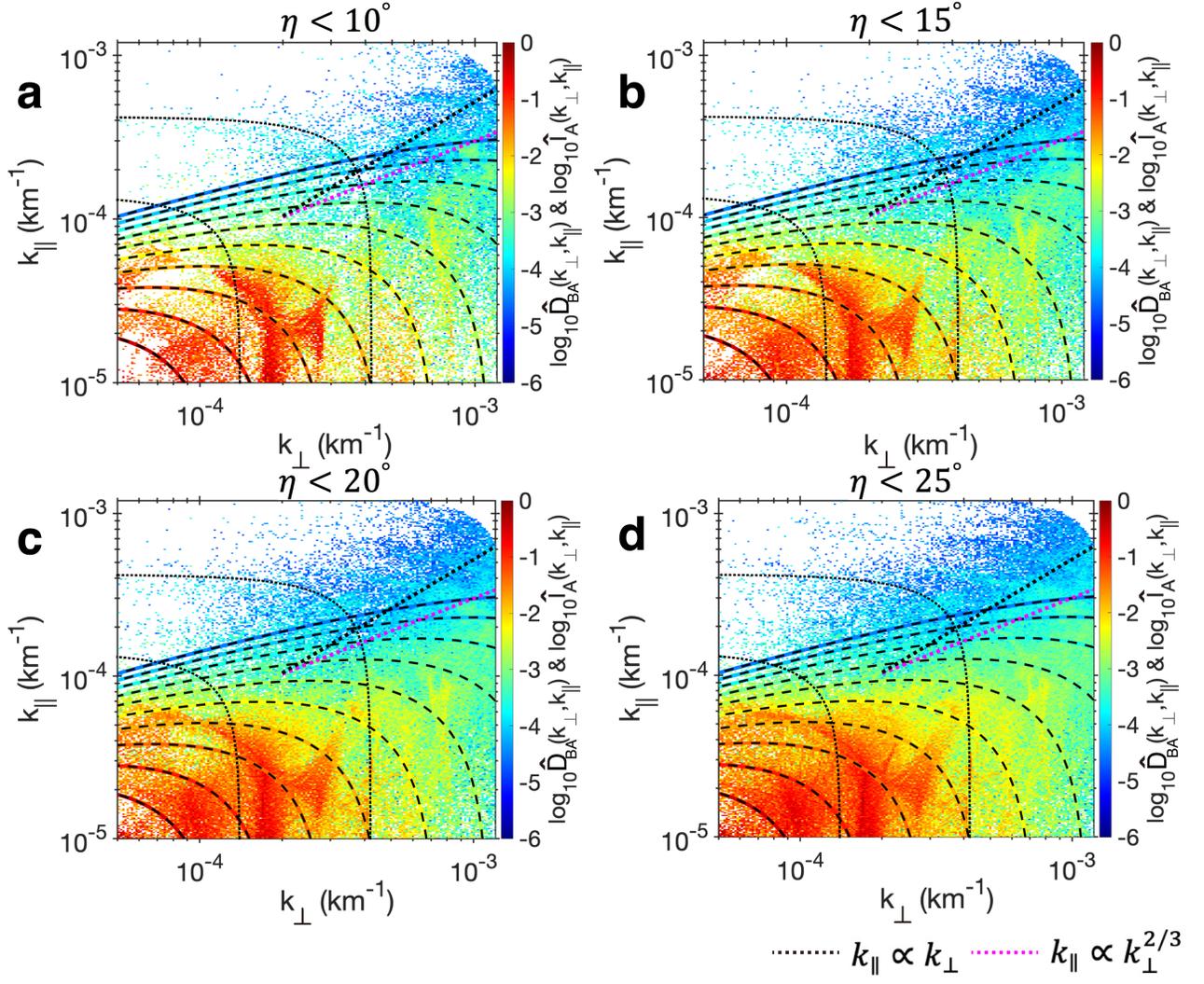}
\caption{2D wavenumber distributions of Alfvénic magnetic energy using data sets under $\eta<10^\circ$, $15^\circ$, $20^\circ$, and $25^\circ$. For each panel, use the same format as Fig. 2a in the main text. The 2D spectral image represents
$\hat{D}_{B_A}(k_\perp,k_\parallel)$. 
The color contours with black dashed curves represent $\hat{I}_{B_A}(k_\perp,k_\parallel)$, which is in the same color map as $\hat{D}_{B_A}(k_\perp,k_\parallel)$. The dotted curves mark $k=\sqrt{k_\parallel^2+k_\perp^2}=0.01/d_i$ and $0.03/d_i$.}
\end{figure}

\setcounter{Sfigure1}{7}
\renewcommand{\thefigure}{Supplementary Fig.\arabic{Sfigure1}}

\begin{figure}
\centering
\includegraphics[width=1\textwidth]{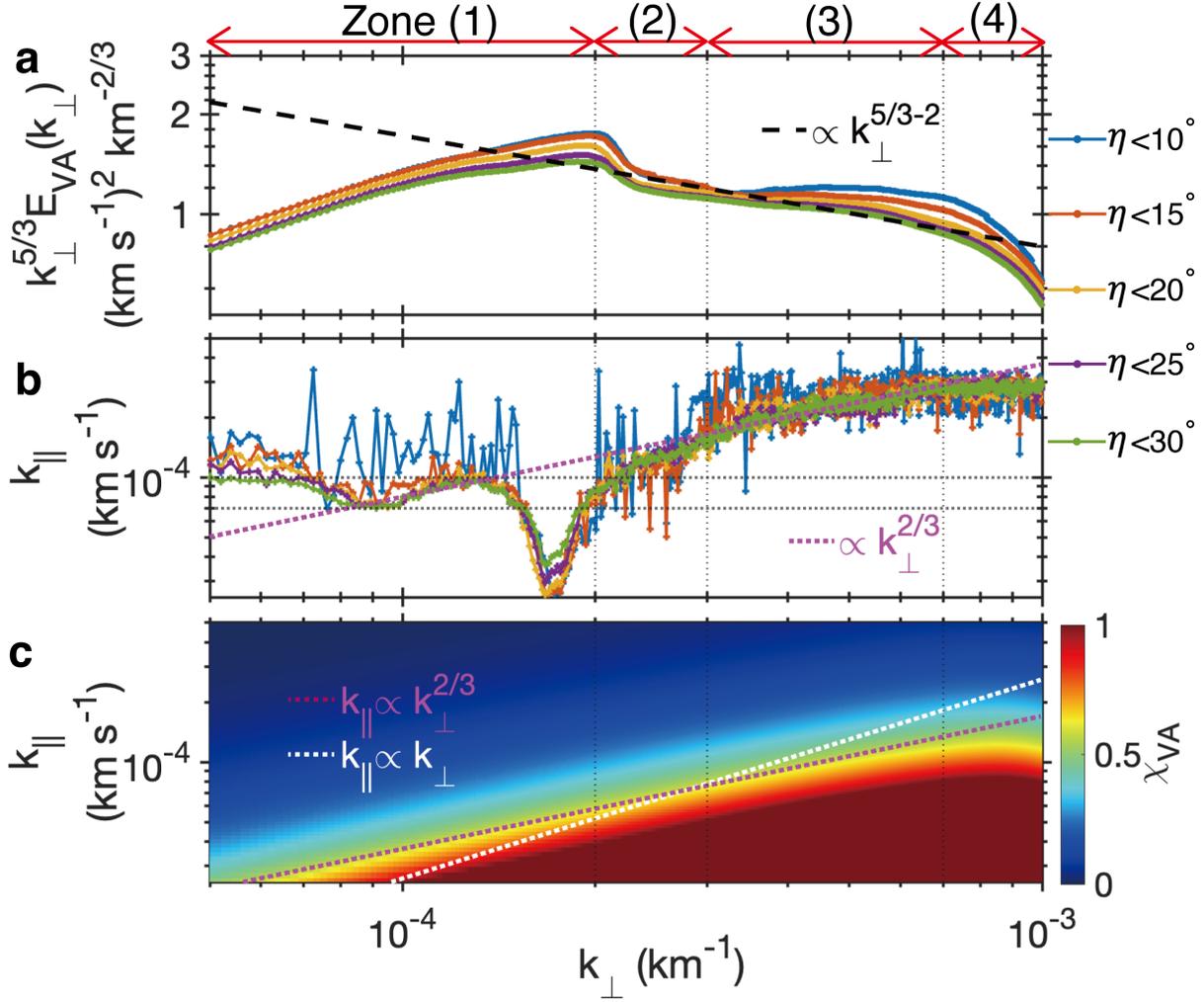}
\caption{Perpendicular wavenumber dependence of the compensated spectra ($k_\perp^{5/3}E_{V_A}(k_\perp)$), parallel wavenumber ($k_\parallel$), and nonlinearity parameter ($\chi_{V_A}(k_\perp,k_\parallel)$). $\mathbf{a}$, $k_\perp^{5/3}E_{V_A}(k_\perp)$ are displayed by curves. The dashed line represents the scaling $k_\perp ^{5/3}E_{V_A}(k_\perp)\propto k_\perp^{5/3-2}$. $\mathbf{b}$, The variation of $k_\parallel$ versus $k_\perp$ by taking the same values of proton velocity energy. The dotted line represents the scaling $k_\parallel \propto k_\perp^{2/3}$. The horizontal dotted lines mark $k_\parallel=7\times10^{-5} km^{-1}$ and $10^{-4} km^{-1}$. $\mathbf{c}$, $\chi_{V_A}(k_\perp,k_\parallel)$ spectrum calculated using the data set under $\eta<30^\circ$. The $k_{\perp}$ dependence figure is divided into four zones: (1) $5\times10^{-5}<k_\perp<2\times10^{-4}km^{-1}$, (2) $2\times10^{-4}<k_\perp<3\times10^{-4}km^{-1}$, (3) $3\times10^{-4}<k_\perp<7\times10^{-4}km^{-1}$, and (4) $7\times10^{-4}<k_\perp<1\times10^{-3}km^{-1}$. The first, second, and third vertical dotted lines are around the maximum of $k_\perp^{5/3}E_{V_A}(k_\perp)$, the beginning and the end of flattened $k_\perp^{5/3}E_{V_A}(k_\perp)$, respectively.}
\end{figure}

\bibliography{sample631}{}
\bibliographystyle{aasjournal}



\end{document}